\documentclass[article]{aa}
\usepackage[varg]{txfonts}

\usepackage{ulem}

\usepackage{hyperref}
\hypersetup{colorlinks=true,linkcolor=blue,citecolor=blue,filecolor=blue,urlcolor=blue}

\usepackage{tikz,xcolor,graphicx}
\graphicspath{{images/}}

\usepackage{bm}

\newcommand{\Lr}{L.r.}
\newcommand{\Msun}{M_\odot}

\newcommand{\trh}[1][]{t_\mathrm{rh#1}}
\newcommand{\tcc}[1][]{t_\mathrm{cc#1}}
\newcommand{\ra}[1][]{r_\mathrm{a#1}}
\newcommand{\rh}[1][]{r_\mathrm{h#1}}
\newcommand{\rt}[1][]{r_\mathrm{t#1}}
\newcommand{\rK}[1][]{r_\mathrm{K#1}}

\newcommand{\disp}[1][]{\sigma_\mathrm{#1}}
\newcommand{\meq}[1][]{m_\mathrm{eq#1}}

\definecolor{lime}{HTML}{A6CE39}
\DeclareRobustCommand{\orcidicon}{%
        \begin{tikzpicture}
        \draw[lime, fill=lime] (0,0) 
        circle [radius=0.16] 
        node[white] {{\fontfamily{qag}\selectfont \tiny ID}};
        \draw[white, fill=white] (-0.0625,0.095) 
        circle [radius=0.007];
        \end{tikzpicture}
        \hspace{-2mm}
}

\newcommand{\orcidVP}{\href{https://orcid.org/0000-0002-3031-062X}{\orcidicon}}
\newcommand{\orcidEV}{\href{https://orcid.org/0000-0003-2742-6872}{\orcidicon}}
\newcommand{\orcidALV}{\href{https://orcid.org/0000-0002-6162-1594}{\orcidicon}}
\newcommand{\orcidDCH}{\href{https://orcid.org/0000-0002-1910-4630}{\orcidicon}}

\defcitealias{pav_ves_letter}{PV1}
\defcitealias{pav_ves2}{PV2}
\defcitealias{pav_ves3}{PV3}

\begin{document}

\title{Dynamics of star clusters with tangentially anisotropic velocity distribution}
\subtitle{}

\titlerunning{Tangential velocity anisotropy in SCs}

\author{V\'aclav Pavl\'ik \inst{\ref{asu},\ref{iu},\ref{ave},}\thanks{\email{pavlik@asu.cas.cz}} \orcidVP
\and Douglas C.~Heggie \inst{\ref{uedin}} \orcidDCH
\and Anna Lisa Varri \inst{\ref{uedin},\ref{ifa}} \orcidALV
\and Enrico Vesperini \inst{\ref{iu}} \orcidEV
}

\institute{
	Astronomical Institute of the Czech Academy of Sciences, Bo\v{c}n\'i~II~1401, 141~00~Prague~4, Czech Republic \label{asu}
	\and Department of Astronomy, Indiana University, Swain Hall West, 727 E 3$^\text{rd}$ Street, Bloomington, IN 47405, USA \label{iu}
	\and Aventinum Publishing House, Tolst\'eho 22, 101~00~Prague~10, Czech Republic \label{ave}
	\and School of Mathematics and Maxwell Institute for Mathematical Sciences, University of Edinburgh, Kings Buildings, Edinburgh EH9 3FD, United Kingdom \label{uedin}
	\and Institute for Astronomy, University of Edinburgh, Royal Observatory, Blackford Hill, Edinburgh, EH9 3HJ, United Kingdom \label{ifa}
}

\authorrunning{Pavl\'ik et al.}

\date{Received: 8 April 2024 / Accepted: 29 May 2024}

\abstract
{Recent high-precision observations with \textit{HST} and \textit{Gaia} enabled new investigations of the internal kinematics of star clusters (SCs) and the dependence of kinematic properties on the stellar mass. These studies raised new questions about the dynamical evolution of self-gravitating stellar systems.}
{We aim to develop a more complete theoretical understanding of how various kinematical properties of stars affect the global dynamical development of their host SCs.}
{We perform $N$-body simulations of globular clusters with isotropic, radially anisotropic and tangentially anisotropic initial velocity distributions. We also study the effect of an external Galactic tidal field.}
{We find three main results.
First, compared to the conventional, isotropic case, the relaxation processes are accelerated in the tangentially anisotropic models and, in agreement with our previous investigations, slower in the radially anisotropic ones. This leads to, e.g., more rapid mass segregation in the central regions of the tangential models or their earlier core collapse.
Second, although all SCs become isotropic in the inner regions after several relaxation times, we observe differences in the anisotropy profile evolution in the outer cluster regions -- all tidally filling models gain tangential anisotropy there while the underfilling models become radially anisotropic.
Third, we observe different rates of evolution towards energy equipartition (EEP). While all SCs evolve towards EEP in their inner regions (regardless of the filling factor), the outer regions of the tangentially anisotropic and isotropic models are evolving to an ``inverted'' EEP (i.e., the high-mass stars having higher velocity dispersion than the low-mass ones). The extent (both spatial and temporal) of this inversion can be attributed to the initial velocity anisotropy -- it grows with increasing tangential anisotropy and decreases as the radial anisotropy rises.}
{}


\maketitle

\section{Introduction}
\label{sec:intro}

Latest observations from the \textit{Hubble Space Telescope} and \textit{Gaia} along with data from large spectroscopic surveys have provided many new insights into the internal kinematical properties of star clusters \citep[SCs; see, e.g.,][]{bellini_hst,watkins_hst,bellini_hstV,bellini_hstII,ferraro_etal,libralato_hst,bianchini_gaia,MUSE,jindal_gaia,cohen_GCs,vasiliev_gaia}.
Complementary studies have shown that the initial velocity distribution of a SC has a significant effect on its dynamical evolution \citep[e.g.,][]{fokkerplanck_rotI,fokkerplanck_rotII,bianchini_meq,cluster_history,imbh_anisotropy,livernois_etal22,LivVesPav23}, and that kinematical signatures are also a useful way of characterising their formation history or the ``multiple stellar population'' conundrum, which remains a focus of current research \citep[e.g.,][]{hst_UV_legacy,hst_legacy_xviii,cordoni_etal20,vesperini_etal21}.
All these findings have highlighted the need for an improved theoretical understanding of SCs, and the recent observational results of \citet{watkins_hst22} further show the importance of studying the different kinematical properties of these stellar systems.

The phenomenological motivation for exploring more complex initial conditions, particularly in the velocity space, is to assess the impact and survivability of any primordial kinematic signatures, ideally, to distinguish them from evolutionary ones.
Therefore, here we present the fourth paper in a series where we investigate the consequences for the stellar kinematics of SCs resulting from dynamical processes in their evolution. In \citet[][hereafter PV1]{pav_ves_letter}, we provided novel insights on the evolution of SCs towards energy equipartition using models with isotropic and radially anisotropic velocity distributions. In \citet[][hereafter PV2]{pav_ves2}, we focused on the strength of the galactic tidal field and the observable parameters to facilitate verification of our results with observations. In \citet[][hereafter PV3]{pav_ves3}, we further studied the properties of binary stars and mass segregation in the isotropic and radially anisotropic SCs\footnote{We note that isotropy, radial and tangential anisotropy in this work always refers to the stellar velocity distributions.}\!\!.

In the current paper, we venture to the other end of the kinematical spectrum, i.e., the lesser known territory of tangential anisotropy.
Anisotropy in the tangential direction has gained some attention in conjunction with equilibrium modelling efforts of stellar systems with a central black hole \citep[see, e.g.,][]{cohn_kulsrud78,young1980,goodman_binney84,feld17,zhang_amaroseoane24}\footnote{It is important to point out that the models in our study do not have a black hole and tangential anisotropy is considered as primordial.}\!\!.
In addition, it was reported to appear in the outskirts of evolved, tidally limited SCs as a result of stellar escapes \citep{Tio_Ves_Var16}, and it was briefly discussed in the context of long-term evolution of SCs by \citet{breen_var_heg} and \citet{rozier_instability}. Recently, \citet{lahen_etal20} gave it some renewed attention after finding signatures of tangential anisotropy in the simulations of low-mass young SC, however, this kinematical regime still remains understudied. Our aim now is to provide more insights into the relaxation processes in tangentially anisotropic SCs and to expand the very few previous works on this topic.

\section{Methods}
\label{sec:methods}

\begin{table}
	\centering
	\caption{Parameters of the models}
	\begin{tabular}{p{1.5cm}cccc}
		\hline\hline
		Model 														& $\beta_0$			& $\ra$			& $\quad f = \rt\,/\,\rK \quad$	& $\tcc\,/\,\trh[,0]$\\
		\hline
		\texttt{iso-u}\,\tablefootmark{a} & $0$						& $\infty$	& $10$	& $3.2$\\
		\texttt{iso-f}\,\tablefootmark{b} & $0$						& $\infty$	& $1$		& $3.0$\\
		\texttt{rh1-u}\,\tablefootmark{a}	& $0$						& $1\,\rh$	& $10$	& $4.6$\\
		\texttt{rh1-f}\,\tablefootmark{b}	& $0$						& $1\,\rh$	& $1$		& $3.8$\\
		\texttt{rad-u}										& $0.5$					& $\infty$	& $10$	& $3.7$\\
		\texttt{rad-f}										& $0.5$					& $\infty$	& $1$		& $3.2$\\
		\texttt{tan-u}										& $-0.5$				& $\infty$	& $10$	& $2.3$\\
		\texttt{tan-f}										& $-0.5$				& $\infty$	& $1$		& $2.4$\\
		\hline
	\end{tabular}
	\tablefoot{
		The columns show: the name, the anisotropy parameters used in Eq.~\eqref{eq:aniso}, the filling factor (filling or underfilling), and the time of core collapse ($\pm0.2\,\trh[,0]$) estimated from the first minimum of the core radius.\\
		\tablefoottext{a}{Models used in \citetalias{pav_ves_letter}.}\\
		\tablefoottext{b}{Models used in \citetalias{pav_ves2}.}
	}
	\label{tab:models}
\end{table}

\begin{figure}
	\includegraphics[width=.9\linewidth]{{{beta_ini_lagr.pdf}}}
	\caption{Initial velocity anisotropy profiles of the models. See also Eq.~\eqref{eq:aniso} and Table~\ref{tab:models}. The radial distance is shown in terms of the initial half-mass radius, each point corresponds to an increment of 2\,\% of mass in the initial Lagrangian radii of the SC.}
	\label{fig:beta_ini}
\vspace{\floatsep}
	\includegraphics[width=.9\linewidth]{{{Trelax_r.pdf}}}
	\caption{Initial relaxation time as a function of radius, calculated from Eq.~\eqref{eq:trelax}.}
	\label{fig:Trel_r}
\end{figure}

\begin{figure*}
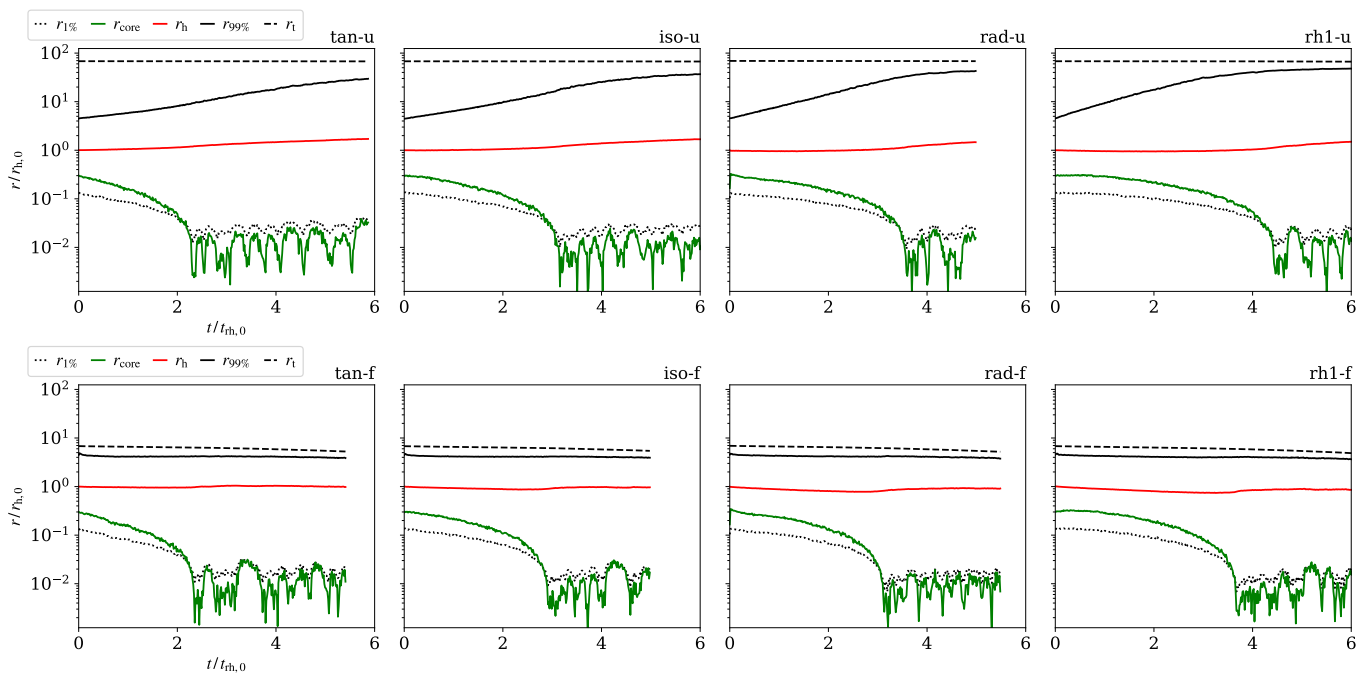

	\includegraphics[width=\linewidth]{{{lagr_0.1.pdf}}} \\[-5pt]
	\includegraphics[width=\linewidth]{{{lagr_1.0.pdf}}}
	\caption{Time evolution of the radial structure of the underfilling SCs (top row) and the filling SCs (bottom row). We show the core radius (as calculated by \textsc{nbody6++gpu}), the 1\,\% and 99\,\% \Lr, the half-mass radius, and the tidal radius (see also the legend).}
	\label{fig:lr-uf}
\end{figure*}

\begin{figure*}[!h]
	\begin{minipage}{\columnwidth}
		\includegraphics[width=.495\linewidth]{{{massloss0.1.pdf}}}
		\includegraphics[width=.495\linewidth]{{{massloss1.0.pdf}}}
		\caption{Total mass of the underfilling (left) and filling models (right) at time $t$, shown as a fraction of the initial SC mass, $M_0$.}
		\label{fig:massloss}
	\end{minipage}
	\hfill
	\begin{minipage}{\columnwidth}
		\includegraphics[width=.495\linewidth]{{{meanmass0.1.pdf}}}
		\includegraphics[width=.495\linewidth]{{{meanmass1.0.pdf}}}
		\caption{Mean stellar mass in Solar masses at time $t$ in the underfilling (left) and filling models (right). The vertical axes have different scales.}
		\label{fig:meanmass}
	\end{minipage}
\end{figure*}

\begin{figure*}[!h]
	\centering
	\includegraphics[width=\linewidth]{{{lagr_LHseparate_0.1.pdf}}} \\[-5pt]
	\includegraphics[width=\linewidth]{{{lagr_LHseparate_1.0.pdf}}}
	\caption{Time evolution of the Lagrangian radii of two stellar mass groups -- low-mass (``L'', $m \leq 0.2\,\Msun$, dotted lines) and high-mass (``H'', $m \geq 0.9\,\Msun$, solid lines). The curves represent 1, 50, 75, and 90 per cent of the total mass of the corresponding mass group (from bottom to top; the half-mass radius is also highlighted with red). The cluster tidal radius is the top dashed line.}
	\label{fig:rlLH}
	\vspace{\floatsep}
	\includegraphics[width=.495\linewidth]{{{segr_gill-etal_0.1.pdf}}}
	\hfill
	\includegraphics[width=.495\linewidth]{{{segr_gill-etal_1.0.pdf}}}
	\caption{Evolution of mass segregation in the SCs. We compare the mean mass in the core and around the half-mass radius (crosses), and around the half-mass radius and in the outer regions (circles). The ranges of Lagrangian radii from which the mean mass is taken are specified as mass percentages in the label. The underfilling models are in the left-hand panel, the filling ones are in the right-hand panel.}
	\label{fig:gill}
\end{figure*}

\begin{figure*}
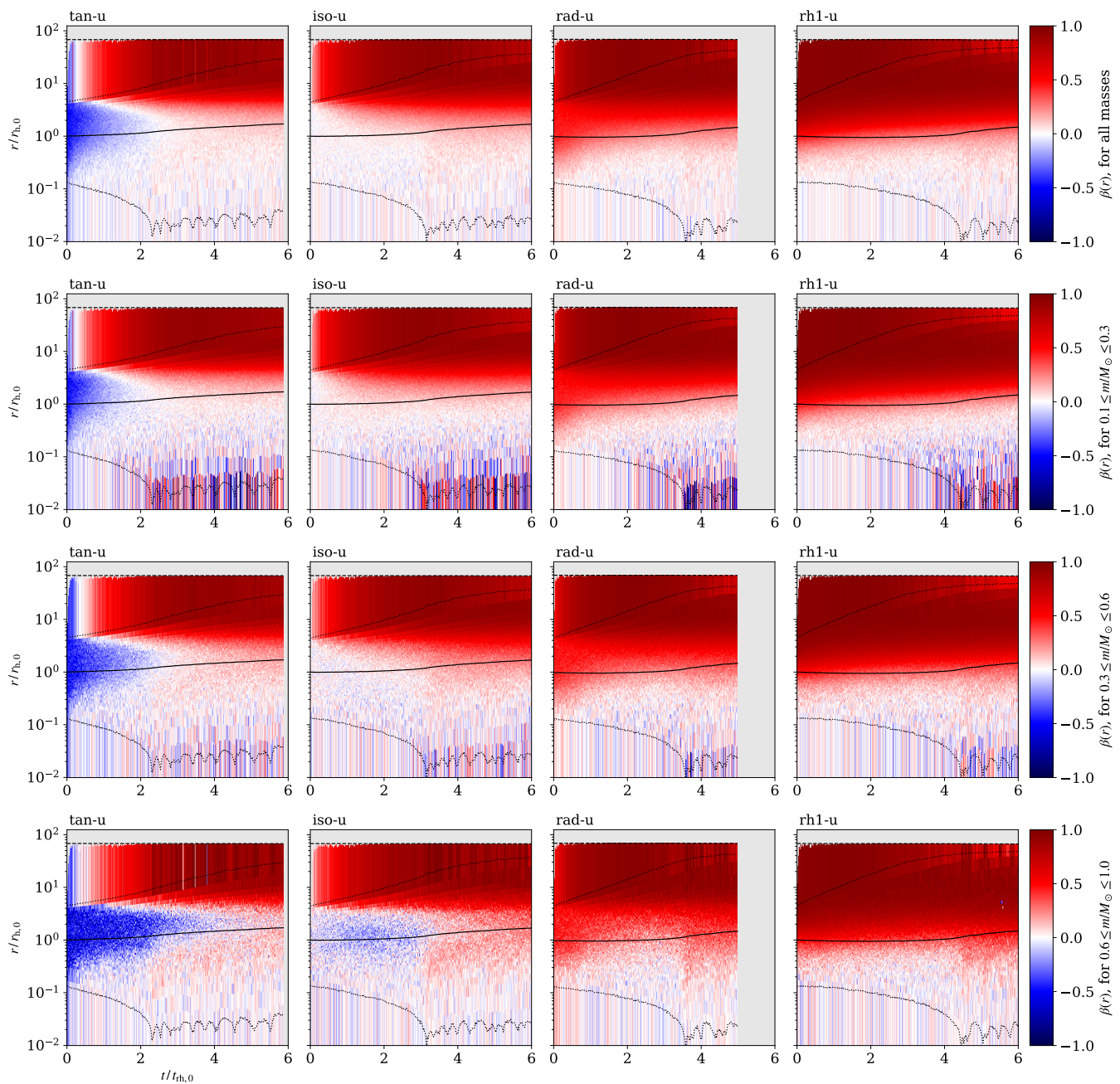

	\includegraphics[width=\linewidth]{{{beta_2p_rm_0.1.pdf}}} \\[-10pt]
	\includegraphics[width=\linewidth]{{{beta_2p_rmL_0.1.pdf}}} \\[-10pt]
	\includegraphics[width=\linewidth]{{{beta_2p_rmM_0.1.pdf}}} \\[-10pt]
	\includegraphics[width=\linewidth]{{{beta_2p_rmH_0.1.pdf}}} \\[-10pt]
	\caption{Time evolution of $\beta(r)$, defined by Eq. \eqref{eq:aniso}, in the underfilling models. The top row shows $\beta$ calculated from the velocity dispersion of all stars in the SC (regardless of mass), the rows below are calculated for specific mass groups -- from the least massive in the second row, intermediate-mass in row three and the most massive stars in the bottom row (see also the labels on the right-hand side). Red colours correspond to radial anisotropy while blue colours to tangential anisotropy. If the fit is poorly defined due to an insufficient number of data points in a radius-mass bin we mask the corresponding pixel with grey colour. To give a sense of the radial scale and dynamical evolution of the SCs in each plot, we also show the 1\,\% \Lr, half-mass radius, 99\,\% \Lr\ and the tidal radius (from bottom to top; black lines). Note that because of the log-scale on the vertical axis, the bins below 2\,\% may appear more prominent than the other bins despite covering less radial range.}
	\label{fig:aniso-u}
\end{figure*}


We simulated $N$-body models with $N_0 = 10^5$ stars, an initial \citet{king_model} profile of $W_0 = 6$, and stellar masses drawn from the \citet{kroupa} initial mass function (IMF) between $0.1 \leq m\,/\,\Msun \leq 1.0$. All models were integrated in \textsc{nbody6++gpu} \citep{nbody6pp}.
There were no primordial binary stars but the dynamical formation of binaries was permitted in the simulations (in the following analyses, those are replaced by their centres of mass).
The modelled clusters were placed on a circular orbit in a Keplerian Galactic potential with two different filling factors -- i.e., the initial ratio of the tidal radius, $\rt$, to the King model truncation radius, $\rK$, was either equal to one or ten. We note that all radii in this work are calculated with respect to the cluster's density centre, as provided by \textsc{nbody6++gpu} \citep[see also][]{casertano_hut,aarseth}.

We considered four sets of initial conditions based on the velocity anisotropy profile. Two of those models -- the isotropic one (\texttt{iso-u/f}) and the one with isotropic distribution of velocities in the core but gradual radial anisotropy (\texttt{rh1-u/f}) -- were already analysed in \citetalias{pav_ves_letter} and \citetalias{pav_ves2}, and are included here to allow for a more direct and detailed comparison. The remaining two models (labelled \texttt{rad-u/f} and \texttt{tan-u/f}) started with a fully radial or fully tangential constant level of anisotropy, respectively. The initial velocity distributions were set up numerically with \textsc{agama}, using the Cuddeford--Osipkov--Merritt profile \citep{osipkov,merritt,cuddeford,agama}
\begin{equation}
	\label{eq:aniso}
	\beta(r) \equiv 1 - \frac{\disp[tan](r)^2}{2 \disp[rad](r)^2} = \frac{\beta_0 + (r/\ra)^2}{1 + (r/\ra)^2} \,,
\end{equation}
where $\disp[rad]$ and $\disp[tan]$ are the radial and tangential stellar velocity dispersions, respectively. The anisotropy radius, $\ra$, determines how steeply does the profile grow towards radial anisotropy. The specific values for each model are listed in Tab.~\ref{tab:models} and the initial radial profiles of $\beta(r)$ are plotted in Fig.~\ref{fig:beta_ini}. We note that the filling and underfilling models start with the same anisotropy profiles, hence the only difference between the two is the size of the tidal radius.

Despite their kinematical differences, all models have the same initial spatial structure and half-mass radii, thus the initial half-mass relaxation time for all of them is given by
\begin{equation}
	\label{eq:trh}
	\trh[,0] = 0.138 \, N_0\,\rh[,0]^{3/2} \,\big/\, \ln{(0.02 N_0)}
\end{equation}
in H\'enon units \citep[see, e.g.,][]{henon_units_1,heggie_hut,henon_units_2}. However, this definition does not take into account the initial velocity distributions which are responsible for different stellar orbits in the SCs and consequently for the variations in the rates of mass-loss, mixing of stars, and the rate of mass segregation. All this leads to different time scales of the relaxation processes (same effect has also been noted by \citealt{breen_var_heg}; \citetalias{pav_ves_letter,pav_ves2,pav_ves3}).

We can see a manifestation of this when we plot the initial relaxation time as a function of radius
\begin{equation}
	\label{eq:trelax}
	t_{\rm relax,0}(r) = \frac{0.34 \, \disp^3(r)}{\langle m (r) \rangle \, \rho(r) \, \ln{\Lambda}}
\end{equation}
in H\'enon units \citep[see][]{heggie_hut,binney_tremaine,merritt_dynamics}.
Here $\Lambda = 0.02 \, N_0$ consistently with Eq.~\eqref{eq:trh},
$\langle m (r) \rangle$ is the mean mass of stars up to the radius $r$,
$\rho(r)$ is their density,
and $\disp(r)$ is their velocity dispersion.
Fig.~\ref{fig:Trel_r} shows that the relaxation time in the core is two to three times lower in the tangentially anisotropic model than the other models and this difference is present approximately up to two half-mass radii.

\section{Results: Underfilling models}
\label{sec:results}

We note that some of the following figures contain both filling and underfilling models for easier visual comparison. Nevertheless, to help us isolate the internal dynamical effects from those related to the external tidal field, we first focus only on the underfilling SCs and compare the filling and underfilling SCs in Sec.~\ref{sec:discuss} below.

\subsection{Structural evolution}
\label{sec:struc-u}

As illustrated in the top row of plots in Fig.~\ref{fig:lr-uf} for the underfilling SCs, the tangential model (\texttt{tan-u}) shows the slowest expansion (see the solid black lines for the 99\,\% Lagrangian radius; hereafter \Lr) and the fastest evolution towards core collapse (see the green line representing the core radius). It is then followed by \texttt{iso-u}, \texttt{rad-u} and \texttt{rh1-u} (with the latter having core collapse more than $2\,\trh[,0]$ later than \texttt{tan-u}). Both dynamical effects (i.e., the rates of expansion and collapse) are caused by the fractions
of tangentially and radially moving stars in the SCs. A larger fraction of stars move within circular shells in the tangentially anisotropic cluster, whereas the radially anisotropic SCs are initialised with a larger number of very eccentric orbits. First, this helps the radially anisotropic SCs to expand more rapidly. Second, stars on radial orbits contribute to the heating-up of the cluster core, delaying its core collapse (see \citetalias{pav_ves_letter,pav_ves2} for further discussion). In the tangentially anisotropic model, on the other hand, the smaller fraction of radially orbiting stars implies that this heat source is reduced. Consequently, the centre of \texttt{tan-u} has smaller central velocity dispersion than the other models and the central relaxation time is smaller (see also Fig.~\ref{fig:Trel_r}).

\subsection{Mass loss and mass segregation}
\label{sec:mass-u}

An important factor which affects the dynamical evolution of the studied systems is mass loss. We visualise it in the left-hand panel of Fig.~\ref{fig:massloss} using the total mass of each SC throughout its evolution. We note that the underfilling models lost at most $5\,\%$ of their initial mass by the end of the simulation -- in particular, \texttt{tan-u} is losing slightly less mass over time than \texttt{iso-u} and a few per cent less than both radially anisotropic SCs. While these numbers are small, they are not negligible when compared to the total mass loss. These differences are a consequence of anisotropy and the number of stars on very elongated eccentric orbits which are more likely to escape.

In the left-hand panel of Fig.~\ref{fig:meanmass}, we include the evolution of the mean stellar mass.
We notice a clear signature of core collapse as the first spike on each curve. Due to the relaxation processes before core collapse, the low-mass stars typically evaporate from the SC at a higher rate than the higher-mass stars (which in turn sink inwards) and the overall $\langle m \rangle$ should increase. We further illustrate this expansion of the low-mass population with the dotted lines in the top row of Fig.~\ref{fig:rlLH}. The radial profiles of low-mass stars are very similar in all underfilling models up to the $75\,\%$ \Lr\ and only differ at higher radii due to a more anisotropy-driven cluster expansion.
Immediately after core collapse, the mean mass is expected to drop down due to the additional ejection of massive stars and hard binaries undergoing close interactions in the central regions, explaining the spike in the left-hand panel of Fig.~\ref{fig:meanmass} \citep[see, e.g., also][]{aarseth1972,fujii_pz,pavl_subr}.

In our case, the left-hand panel of Fig.~\ref{fig:meanmass} further reveals that both radially anisotropic models (\texttt{rad-u} and \texttt{rh1-u}) exhibit the highest increase of the mean mass and this growth continues even after core collapse. This is due to the higher evaporation of stars on extremely eccentric orbits.
On the other hand, the mean mass in the tangentially anisotropic SC (\texttt{tan-u}) stays almost constant even before core collapse, which is mainly caused by the very low initial mass loss (see again the yellow dashed line in the left-hand panel of Fig.~\ref{fig:massloss}). Unlike the radial models, \texttt{iso-u} and \texttt{tan-u} also show almost no evolution of $\langle m \rangle$ in the post-core-collapse phase.

As for mass segregation, \citetalias{pav_ves3} previously concluded that it proceeds at different rates in the outer and inner regions of isotropic and radially anisotropic SCs. Their models were similar to our \texttt{iso-u} and \texttt{rh1-u} but contained 10\,\% primordial binaries which we have not included in the simulations presented in this paper. Nonetheless, the same conclusion can be drawn from our models as well. Specifically, in the top row of Fig.~\ref{fig:rlLH}, the lowermost solid line representing the 1\,\% \Lr\ of the massive stars has slower contraction in both radially anisotropic models than in the isotropic or tangentially anisotropic one. On the contrary, the uppermost solid line, which shows the 90\,\% \Lr, significantly decreases in both radial models (\texttt{rh1-u} and \texttt{rad-u}), only slowly decreases in \texttt{iso-u} but stays constant in the tangentially anisotropic one (see the top-left panel of Fig.~\ref{fig:rlLH}).

We further investigate the rate of mass segregation by adopting the parameter introduced by \citet{gill_etal} in the left-hand panel of Fig.~\ref{fig:gill} where we compare the mean stellar masses in three regions, pairwise -- the core (0--5\,\% \Lr), around the half-mass radius (45--50\,\% \Lr), and the outer regions (70--75\,\% \Lr). There is a clear indication that before core collapse, \texttt{tan-u} experiences the slowest mass segregation in the outer regions out of all models (see how the yellow circles are below the other circles). However, it has somewhat more rapid mass segregation than the other models in the inner regions initially (see the yellow crosses on top of all the datasets).
As discussed in \citetalias{pav_ves3}, the speeding-up of mass segregation in the outer regions of the anisotropic model (relative to the segregation rate in the isotropic model) was caused by the enhanced rate of migration of the massive stars on radial orbits, and the delay in mass segregation in the core of the same model was due to the heating-up by the larger number of stars on eccentric orbits. We can extrapolate their argument even to the models in our work -- specifically, the tangentially anisotropic one (\texttt{tan-u}), which has very few of these elongated orbits, has the shortest relaxation time in the core and the fastest mass segregation there. The other models are then almost ordered in the left-hand panel of Fig.~\ref{fig:gill} by the amount of initial radial anisotropy (we note that although \texttt{rad-u} has a constant level of radial anisotropy at all $r$, \texttt{rh1-u} reaches higher values of $\beta$ in the outer regions which makes it more radially anisotropic in this sense).

\subsection{Anisotropy evolution}
\label{sec:aniso-u}

In this section, we offer a more detailed look on the internal kinematics in the underfilling models. Starting from the initial $\beta(r)$ profile, displayed in Fig.~\ref{fig:beta_ini}, here we plot its temporal evolution in Fig.~\ref{fig:aniso-u} (see also Appendix~\ref{app:meq} for technical details on how these plots were made).
To explore the possible dependence of anisotropy on the stellar mass, in addition to the overall $\beta(r)$ profile (in the top row of Fig.~\ref{fig:aniso-u}), we also show the anisotropy evolution in three distinct mass groups (as labelled in each of the lower rows of the figure).
Two major global features are apparent in the top row of Fig.~\ref{fig:aniso-u} for all models. First, the regions below the half-mass radius (which is represented by the solid black middle line in each plot) are mostly isotropic (plotted with white) -- since the relaxation time decreases by one to two orders of magnitude towards the SC core (as we visualised in Fig.~\ref{fig:Trel_r}), the initial anisotropy of the velocity distribution can be quickly removed. Second, the outer regions have significant radial anisotropy (see the red area in the figure) -- this is consistent with the findings of \citet{Tio_Ves_Var16} and is due to the ongoing relaxation processes as well as the overall expansion of the SC, which we have discussed above in Sec.~\ref{sec:mass-u}.

The only model that displays tangential anisotropy is \texttt{tan-u}, which preserves its initial tangential anisotropy imprint during the first $1{-}2\,\trh[,0]$ (see the blue region in the top left plot of Fig.~\ref{fig:aniso-u}). However, similarly to the other models, it also becomes isotropic in the inner regions after several relaxation times. A~comparison of the blue wedges in the left-most panels in all rows of Fig.~\ref{fig:aniso-u} further reveals that higher-mass stars in \texttt{tan-u} maintain their tangential anisotropy around and above the half-mass radius for longer than the lower-mass stars. This helps us explain the previously discussed results on the levels of mass segregation (i.e., that it proceeds at the slowest rate in the outer regions of \texttt{tan-u}). If we focus on the other panels in the figure, we can also notice that no SC develops tangential anisotropy on its own. The only exception is the high-mass population of \texttt{iso-u} where
mild tangential anisotropy appears before core collapse around the half-mass radius (see the faint blue spot around the half-mass radius line in the bottom row of Fig.~\ref{fig:aniso-u}). This is because the massive stars on more radial orbits tend to be mass-segregating into the core, leaving behind the ones on more tangential orbits.

After core collapse, the most massive stars, which segregated to the core, are being ejected from the central region. This dynamical activity causes high fluctuations of $\beta$ between positive and negative values around 1\,\% \Lr, and increases the radial anisotropy in the inner parts of the SCs, which is visible in the bottom row of Fig.~\ref{fig:aniso-u} as the pale red area just below the half-mass radius. Consistently with our previous discussion about the degree of mass segregation, the tangential model shows an overall smallest increase in $\beta$, followed by \texttt{iso-u} and then the two radially anisotropic models.

\begin{figure*}
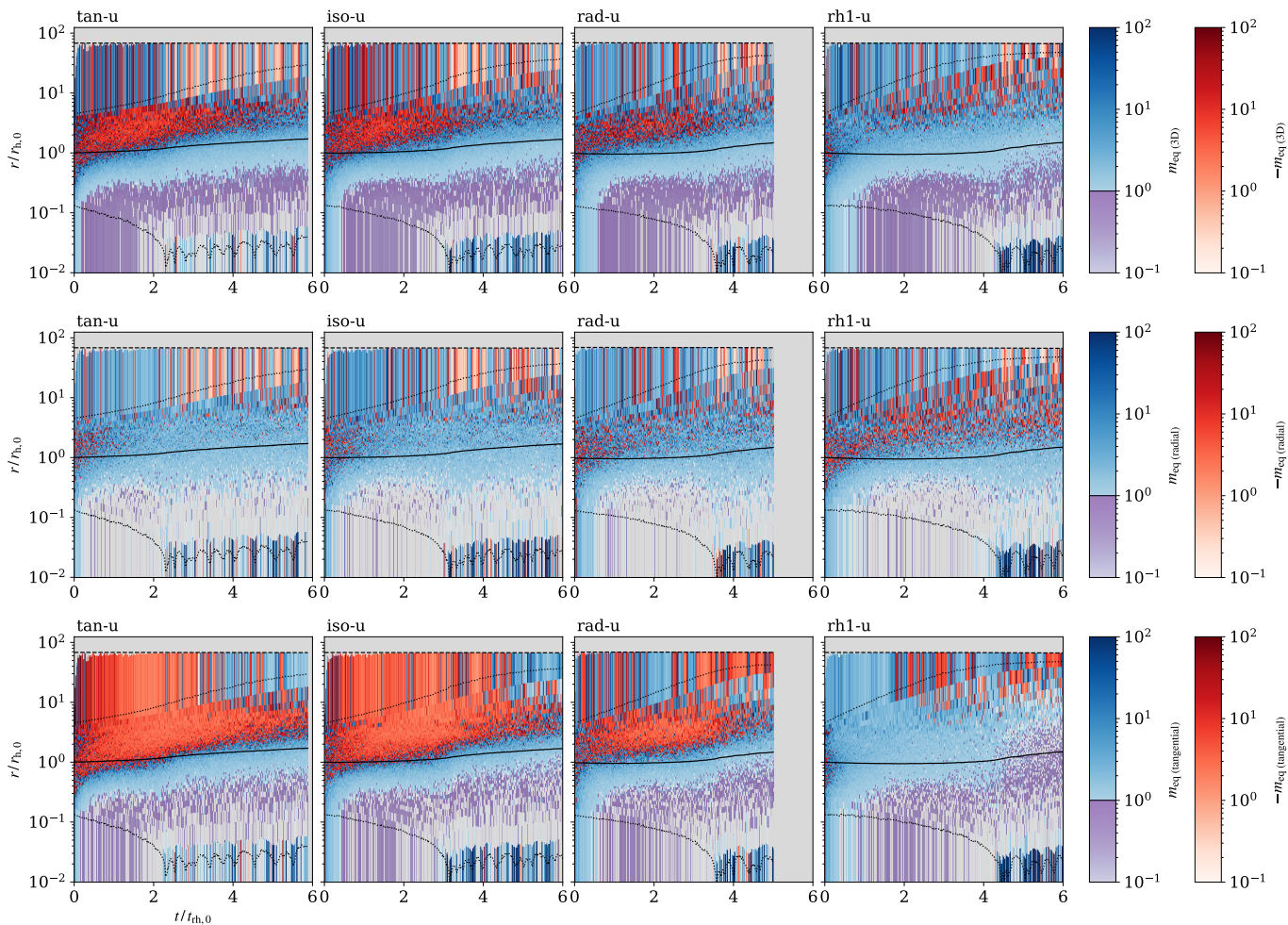

	\includegraphics[width=\linewidth]{{{meq_2p_0.1-fit.pdf}}} \\[-10pt]
	\includegraphics[width=\linewidth]{{{meqR_2p_0.1-fit.pdf}}} \\[-10pt]
	\includegraphics[width=\linewidth]{{{meqT_2p_0.1-fit.pdf}}} \\[-10pt]
	\caption{Time evolution of the equipartition mass, $\meq$, in the underfilling SCs (top row -- calculated from the three-dimensional velocity dispersion; middle row -- calculated only from its radial component; bottom row -- calculated only from its tangential component). The regions which evolve towards EEP (to lower positive values of $\meq$) are in shades of blue and purple, the regions which evolve away from EEP (to negative $\meq$, i.e., away from energy equipartition) are in shades of red, see the colour bars. To give a sense of the radial scale and dynamical evolution of the SCs in each plot, we also show the 1\,\% \Lr, half-mass radius, 99\,\% \Lr\ and the tidal radius (from bottom to top; black lines). If the fit is poorly defined due to an insufficient number of data points in a radius-mass bin we mask the corresponding pixel with grey colour. Note that because of the log-scale on the vertical axis, the bins below 2\,\% may appear more prominent than the other bins despite covering less radial range.}
	\label{fig:meq-u}
\end{figure*}

\begin{figure*}[!h]
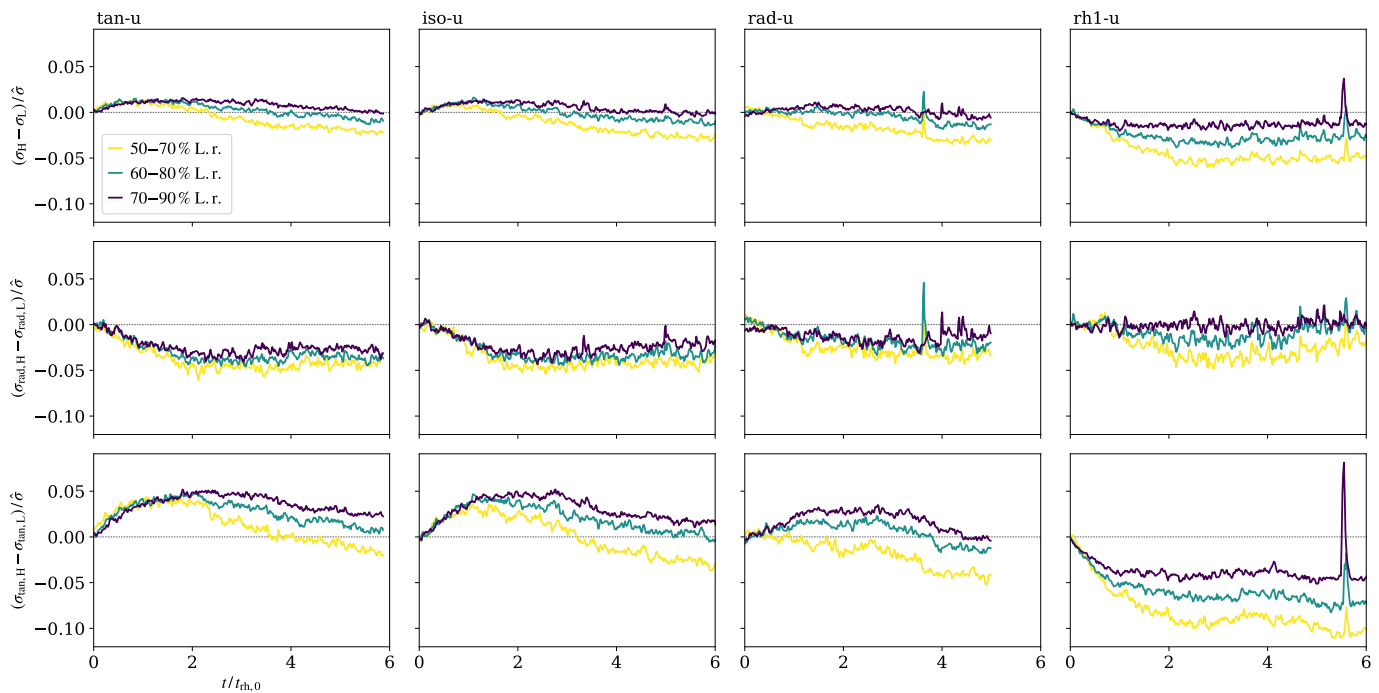

	\includegraphics[width=\linewidth]{{{new_disp_LH_0.1highres.pdf}}}
	\caption{Time evolution of the difference between the velocity dispersion of the high-mass stars (`H', $0.8 \leq m/\Msun \leq 1.0$) and the low-mass stars (`L', $0.1 \leq m/\Msun \leq 0.2$). The velocity dispersion is scaled by $\hat{\disp} = \rh[,0]^{-1/2}$. The rows show the difference in the total $\disp$ (top), its radial component (middle) and its tangential component (bottom). The columns are separated by models (as labelled on top). Each plot has three lines which represent three outside regions of the clusters, delimited by the listed Lagrangian radii.}
	\label{fig:disp-u}
\end{figure*}

\subsection{Energy equipartition}
\label{sec:meq-u}

As a SC evolves, the random encounters between stars of different masses guide the system towards energy equipartition \citep[hereafter EEP; see also, e.g.,][]{spitzer69,inagaki_wiyanto,webb_vesperini_b}. We can quantify the ``degree'' of EEP with the equipartition mass, $\meq$ (a parameter introduced by \citealt{bianchini_meq}, extended to allow for positive and negative values by \citetalias{pav_ves_letter} and formally redefined by \citealt{aros_meq}):
\begin{equation}
	\label{eq:meq}
	\disp(r,m) = \left\lbrace
		\begin{array}{ll}
			\disp[0] \exp{[-m / (2\meq)]} & \text{, } \meq < 0 \\
			\disp[0] \exp{[-m / (2\meq)]} & \text{, } \meq \geq 0 \text{ and } m \leq \meq \\
			\disp[0] e^{-0.5} \sqrt{\meq / m} & \text{, } \meq \geq 0 \text{ and } m > \meq
		\end{array}
		\right.
\end{equation}
Here $m$ is the stellar mass and $\disp[0]$ is a scaling parameter (the velocity dispersion at zero mass). Infinite or very high $|\meq|$ shows that there is no relationship between the stellar mass and velocity (i.e., no EEP), whereas if $\meq$ decreases below some stellar mass $m$ and $\meq > 0$, particles above that mass achieved EEP. Despite the expectations for evolved systems, some SCs do not reach full EEP \citep[see, e.g.,][]{spitzer69,vishniac,omegaCen_noequip}. In \citetalias{pav_ves_letter}, we also found that several SCs develop a trend between velocity dispersion and mass in their outer regions which is opposite to EEP, i.e., low-mass stars have systematically lower $\disp$ than high-mass stars. Here, we extend these previous results.

In Fig.~\ref{fig:meq-u}, we plot the evolution of the equipartition mass in the underfilling models. In essence, these plots are similar to those in \citetalias{pav_ves_letter,pav_ves2,pav_ves3}, however, here we now include two new models and use a much higher radial resolution (compared to only three radial bins in the former works).

In the top row of Fig.~\ref{fig:meq-u}, we can see that the SCs evolve towards EEP in the inner regions up to ${\lesssim}20\,\%$ \Lr\, However, only the most massive stars $m \gtrsim 0.5\,\Msun$ show long-term signs that they achieved EEP (see the purple-coloured bins before and around $2\,\trh[,0]$). In certain very central bins, the lowest value of $\meq$ occasionally drops as low as $0.4\,\Msun \pm 0.1\,\Msun$ (the tangential and radial components of $\disp$ may give even slightly lower values of $\meq$, however, with higher uncertainties), but it never stays this low for more than one time snapshot. The full mass spectrum is, therefore, never in EEP in our SCs. This behaviour is similar for all models.
On the other hand, the outer regions differ despite having similar levels of anisotropy as we saw in Sec.~\ref{sec:aniso-u}. The only underfilling model that has some tendency to evolve towards EEP in the region above the half-mass radius is \texttt{rh1-u} (see the top-right plot in Fig.~\ref{fig:meq-u}) but $\meq$ still fluctuates between positive and negative values (switching between blue and red colours). Thus, we can also draw the conclusion that the system remains in its initial state of velocity equipartition in which the velocity dispersion is independent of the stellar mass (i.e., the slope of $\disp(m)$ is near flat and $\meq$ oscillates around $\pm\infty$).

The radial component of the velocity dispersion rarely contributes to EEP, as we can see from the scattered purple bins in the middle row of Fig.~\ref{fig:meq-u}. The tangential component, however, contributes to EEP the most (see the bottom row of Fig.~\ref{fig:meq-u}). Moreover, in the post-core-collapse evolution of, e.g., the \texttt{rh1-u} model, the regions where $\disp[tan]$ shows EEP among the high-mass stars extend up to the half-mass radius and the outer regions (see the purple bins in the bottom right panel).

Fig.~\ref{fig:meq-u} further shows that \texttt{rad-u}, \texttt{iso-u} and \texttt{tan-u} develop an opposite velocity-mass distribution to what would be expected for a system in partial or full EEP (see the red regions above $\rh$ in the top row of the figure). In \texttt{rad-u} and \texttt{iso-u} models, it takes approximately until the moment of core collapse for this trend to disappear; then the outer regions start showing high fluctuations between positive and negative values, as in the case of \texttt{rh1-u}.
For the model \texttt{tan-u}, the red region influenced by the negative $\meq$ extends up to several relaxation times after core collapse and is also the most extended (reaching down to ${\approx}0.5\,\rh$, initially). In comparison, this region is less extended in the \texttt{iso-u} and \texttt{rad-u} models (see the middle two plots in the top row of Fig.~\ref{fig:meq-u}).

As illustrated in the bottom two rows of Fig.~\ref{fig:meq-u}, where we plot $\meq$ calculated from the radial and tangential components of $\disp$, respectively, the systematically negative $\meq$ (red colour) is clearly visible only in the tangential component. On the other hand, in the radial component, if there is $\meq<0$, it is rather a fluctuation between positive and negative values (i.e., around $\meq \rightarrow \pm\infty$).
This is consistent with the discussion in \citetalias{pav_ves2}, stating that the inversion of EEP (i.e., the high-mass stars having higher velocity dispersion than the low-mass ones) is mainly apparent in the tangential component of velocity dispersion.

We further corroborate this result with all our models in Fig.~\ref{fig:disp-u} where we plot the difference between the velocity dispersion of the high-mass ($\disp[H]$) and the low-mass stars ($\disp[L]$). The top row of plots with the total velocity dispersion shows higher $\disp[H]$ than $\disp[L]$ in the whole region above the half-mass radius in the models \texttt{tan-u} and \texttt{iso-u} up to core collapse; and similarly in the \texttt{rad-u} model but only in the 70--90\,\%\,\Lr\ region. The bottom row of Fig.~\ref{fig:disp-u} shows that this observed behaviour is due to the tangential component of $\disp$.

The reason behind the observed behaviour of $\disp$ and $\meq$ in the outer regions is as follows. As we briefly mentioned in Sec.~\ref{sec:aniso-u}, the more massive stars at large radius may be thought of as two populations: those with more tangential motion (which stay out of the central, dense regions, do not relax much, are subject to little mass segregation, and so remain in the outer regions for a long time) and those with more radial motions (which therefore visit the central, dense regions, relax more quickly, are subject to greater mass segregation, and so are removed from the outer regions).
Thus, tangential motions of high-mass stars in the outer regions tend to stay higher in \texttt{tan-u} than \texttt{iso-u} or the other models because of the initially lower removal rate of stars with low tangential motions.
As discussed in Sec.~\ref{sec:mass-u}, the low-mass stars migrate outwards from the inner regions and their motion is more in the radial direction than the tangential direction. Consequently, this decreases their tangential velocity dispersion in the outer regions (see also \citetalias{pav_ves_letter} and \citetalias{pav_ves2} for further discussion of the effect of migration of the low-mass stars).

The net result of all this is a positive difference between the velocity dispersion of the high-mass and the low-mass stars in the outer regions (i.e., opposite to what is expected for systems evolving towards EEP).
As the relaxation processes in the outer regions become more and more prominent, $\disp[H]$ also starts to decrease and the initial positive difference between $\disp[H]$ and $\disp[L]$ gradually vanishes (see how the curves representing the different \Lr\ shells in Fig.~\ref{fig:disp-u} become ordered by the mass percentages in all models over time).

Furthermore, the ratio of tangentially to radially moving stars above the half-mass radius seems to directly affect the sign of $\meq$ calculated from the tangential component of $\disp$ in this region. In other words, the lower the initial value of $\beta$ is, the longer the SC tends to remain in the negative $\meq$ regime. The model \texttt{rh1-u} is an extreme case where the outer region is almost fully radially anisotropic initially ($\beta \approx 1$) and $\meq[\,\rm{(tangential)}]$ is basically never negative (although it has the tendency to display negative $\meq[\,\rm{(radial)}]$ -- compare the plots in the right-hand column of Fig.~\ref{fig:meq-u}). On the other hand, the model \texttt{tan-u} has $\beta \approx -0.5$ and is, therefore, the other extreme where we see the longest duration of $\meq<0$ in the outer regions (compare the plots in the left-hand column of Fig.~\ref{fig:meq-u}). The other two models fit in-between these extremes based on their initial $\beta(r)$ (see the middle two columns of Fig.~\ref{fig:meq-u}).

\section{Results: Filling vs.\ underfilling models}
\label{sec:discuss}

\begin{figure*}
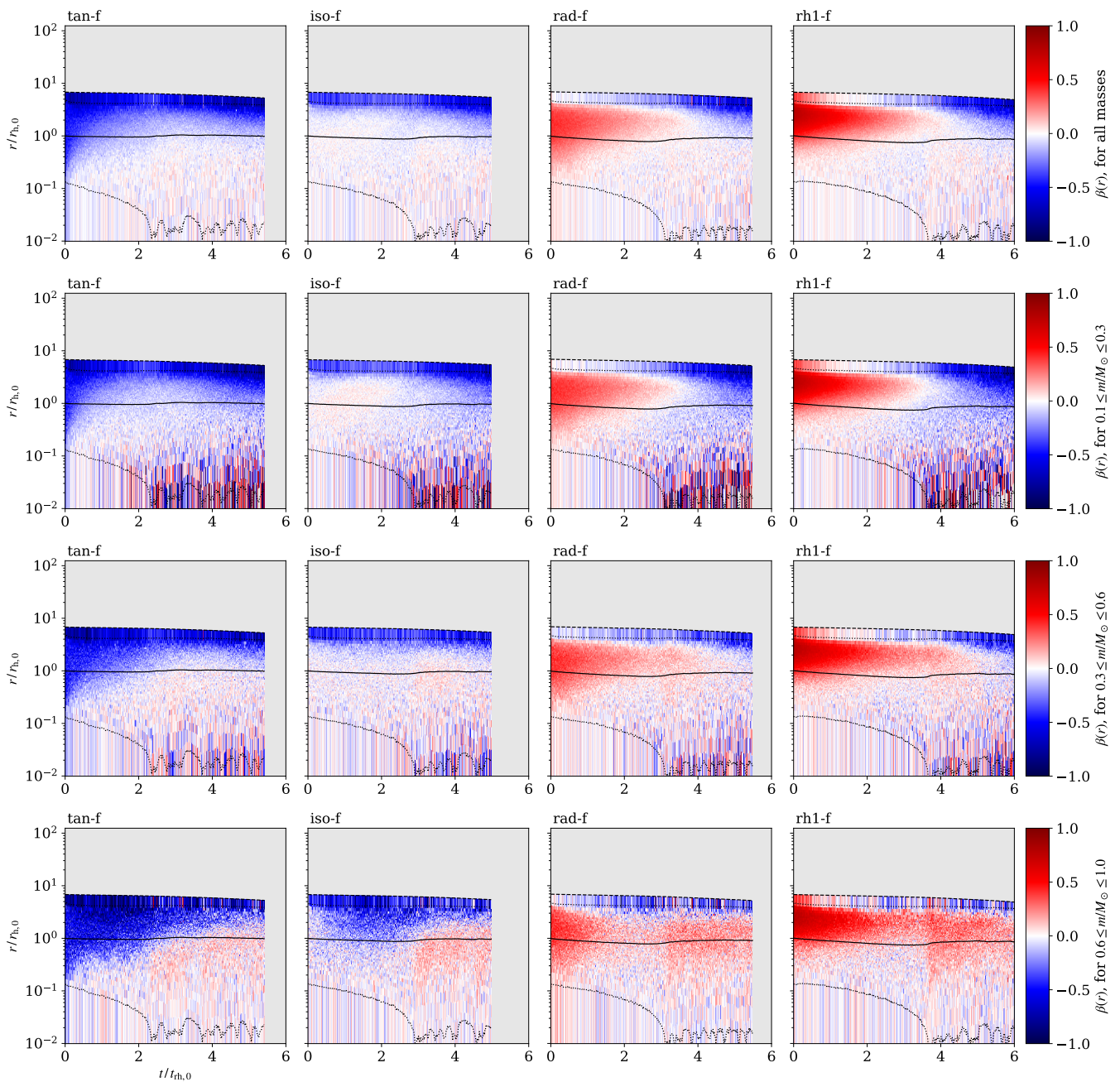

	\includegraphics[width=\linewidth]{{{beta_2p_rm_1.0.pdf}}} \\[-10pt]
	\includegraphics[width=\linewidth]{{{beta_2p_rmL_1.0.pdf}}} \\[-10pt]
	\includegraphics[width=\linewidth]{{{beta_2p_rmM_1.0.pdf}}} \\[-10pt]
	\includegraphics[width=\linewidth]{{{beta_2p_rmH_1.0.pdf}}} \\[-10pt]
	\caption{Same as Fig.~\ref{fig:aniso-u} but for the filling models. The vertical axis has the same range as in Fig.~\ref{fig:aniso-u} to ease the comparison; we note, however, that the tidal radius limits the SC at a much lower $r$.}
	\label{fig:aniso-f}
\end{figure*}

\begin{figure*}[!h]
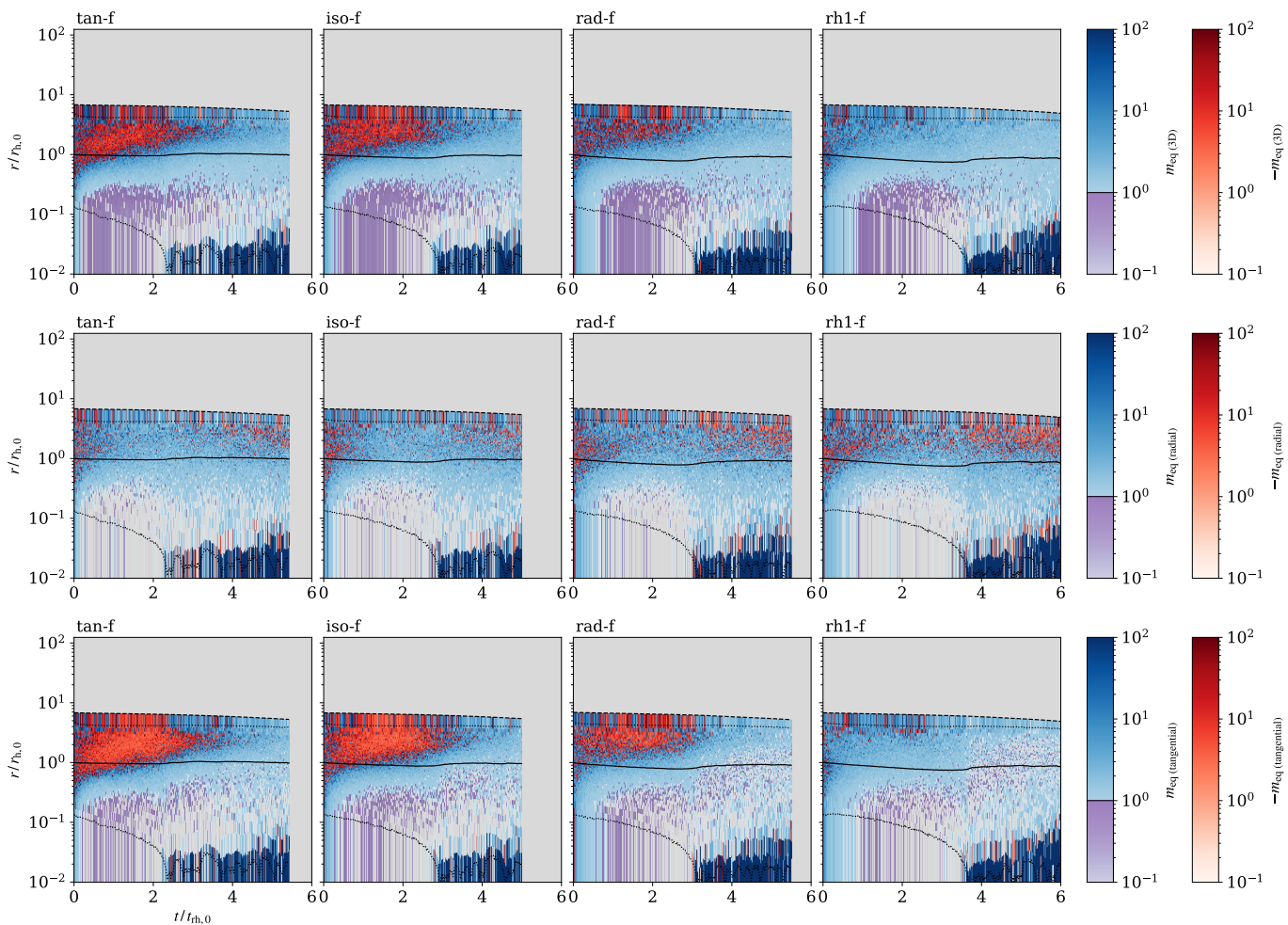

	\includegraphics[width=\linewidth]{{{meq_2p_1.0-fit.pdf}}} \\[-10pt]
	\includegraphics[width=\linewidth]{{{meqR_2p_1.0-fit.pdf}}} \\[-10pt]
	\includegraphics[width=\linewidth]{{{meqT_2p_1.0-fit.pdf}}} \\[-10pt]
	\caption{Same as Fig.~\ref{fig:meq-u} but for the filling models. The vertical axis has the same range as in Fig.~\ref{fig:meq-u} to ease the comparison; we note, however, that the tidal radius limits the SC at a much lower $r$.}
	\label{fig:meq-f}
\end{figure*}

In this section, we discuss the models that started as tidally limited and compare them to their underfilling counterparts.

\subsection{Structural evolution}
\label{sec:struc-f}

In Sec.~\ref{sec:struc-u}, we pointed out a trend concerning the order in which the SCs experience core collapse. As Table~\ref{tab:models} and the evolution of the \Lr\ (bottom row of Fig.~\ref{fig:lr-uf}) indicate, this trend seems to be independent of the filling factor. However, since the tidally limited SCs have no room to expand, the stars on the most eccentric orbits with large semi-major axes are being removed more efficiently. Consequently, the role of radial anisotropy in delaying core collapse is reduced and the difference in $\tcc$ is not as prominent as in the underfilling models -- specifically, \texttt{iso-f} and \texttt{rad-f} are only separated by ${\approx}0.2\,\trh[,0]$ (which is within the estimated uncertainties), and the largest span between \texttt{tan-f} and \texttt{rh1-f} is only about $1.4\,\trh[,0]$ (compared to ${\gtrsim}2\,\trh[,0]$ in the underfilling counterparts).

\subsection{Mass loss and mass segregation}
\label{sec:mass-f}

We quantify mass loss in the filling models in the right-hand panel of Fig.~\ref{fig:massloss} by plotting the total SC mass in time.
Note that while the underfilling SCs only lost a few percent of their mass over the course of several relaxation times, the filling models lost around $50{-}60\,\%$ by the end of the simulations. The initial slope of the mass-loss curves is also different. There is almost no initial mass loss in the underfilling models but all the filling ones show a short but steep initial decrease in the total mass, caused by the primordial escapers. Furthermore, for the underfilling models, there is a change in the rate of mass loss at core collapse (see the break in the slopes of the dashed lines in the inset plot in the left-hand panel of Fig.~\ref{fig:massloss}) while for the tidally filling clusters the mass loss proceeds at almost the same rate, regardless of the initial kinematics and the evolutionary stage. This shows that the tidal limitation is a much bigger factor in the overall mass loss than the initial anisotropy in the filling models.

The increased mass loss also affects the evolution of the mean stellar mass, plotted in the right-hand panel of Fig.~\ref{fig:meanmass}. It shows a similar growth in all the filling SCs -- with both radially anisotropic SCs increasing only slightly faster (by ${\approx}0.0025\,\Msun/\trh[,0]$) than the isotropic or tangentially anisotropic ones. There is also no visible signature of core collapse in the mean stellar mass of the filling models.

As for mass segregation in the filling models, the dotted lines in the bottom row of Fig.~\ref{fig:rlLH} show that the radial profiles of the low-mass stars are almost identical throughout the SCs evolution (which also agrees with the similarities in the evolution of the mean mass and the rates of mass loss). The Lagrangian radii of the high-mass stars (solid lines in the bottom row of the same figure) follow a qualitatively similar trend as in the underfilling models (top row of the figure). Furthermore, we find similar degrees of mass segregation in the core of both the filling and the underfilling models, just on a slightly different timescales (compare the crosses between the right and left-hand panels of Fig.~\ref{fig:gill}). The outer regions (circles in the same figure) also evolve in a similar way, however, due to the high ongoing mass loss in the filling models, its outer $\langle m \rangle$ is growing even after core collapse.

\subsection{Anisotropy evolution}
\label{sec:aniso-f}

The obvious difference between the evolution of the $\beta(r)$ profile in the filling (Fig.~\ref{fig:aniso-f}) and underfilling SCs (Fig.~\ref{fig:aniso-u}) is in the outer regions. As the majority of radially moving stars is removed from the filling clusters, more of those on tangential orbits remain. Consequently, the SCs develop an outer shell with significant tangential anisotropy -- see the blue regions around the tidal radius in Fig.~\ref{fig:aniso-f} \citep[this is, again, consistent with][]{Tio_Ves_Var16}. This feature is independent of the stellar mass.

The regions below the half-mass radius show no qualitative differences between the filling and underfilling SCs before core collapse and only minor deviations can be found after $\tcc$. Specifically, the low-mass stars in the filling models develop more tangential anisotropy in the shell from $0.1\,\rh[,0]$ to $1\,\rh[,0]$, whereas the high-mass and intermediate-mass stars are slightly less radially anisotropic around the half-mass radius. The whole core region is again noisy, therefore, we do not discuss it here.

\subsection{Energy equipartition}
\label{sec:meq-f}

\begin{figure*}[!h]
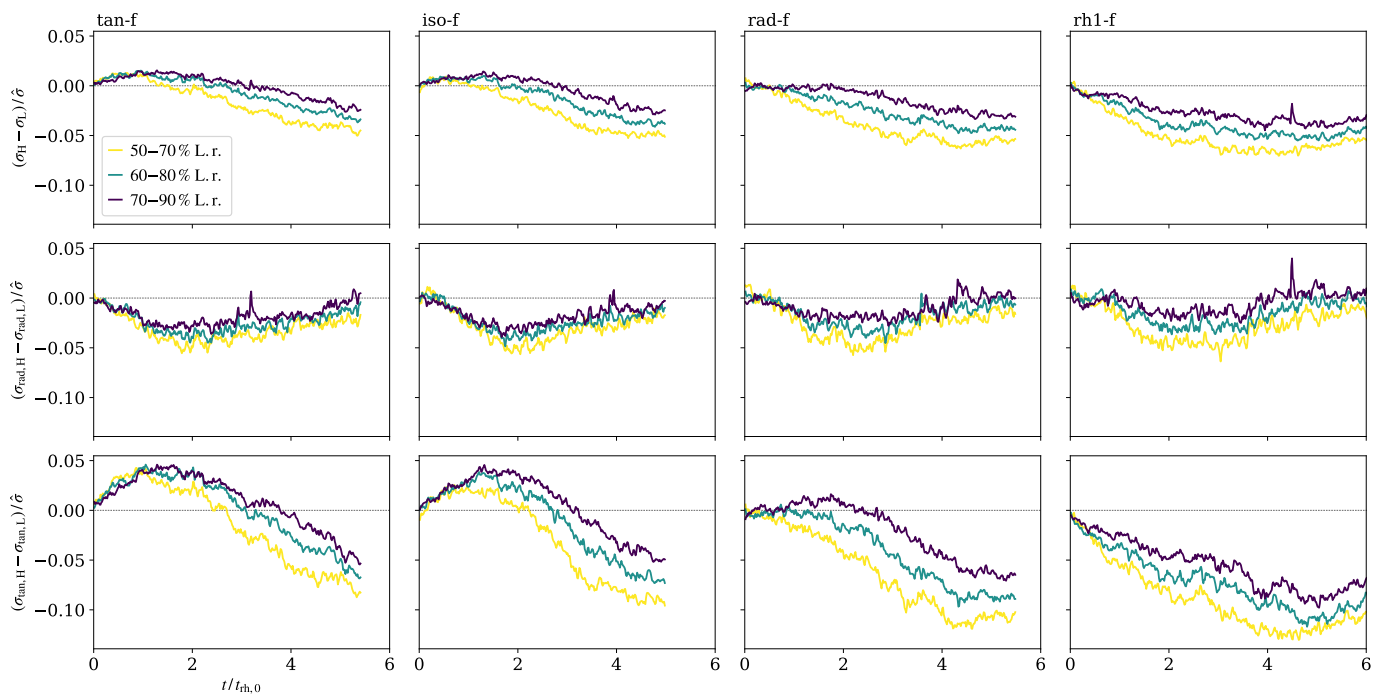

	\includegraphics[width=\linewidth]{{{new_disp_LH_1.0highres.pdf}}}
	\caption{Same as Fig.~\ref{fig:disp-u} but for the filling models.}
	\label{fig:disp-f}
\end{figure*}

We also find similarities when studying the equipartition mass. Apart from the different timescales, the regions up to the half-mass radius look almost identical in the underfilling models and their filling counterparts (see Figs.~\ref{fig:meq-u} \&~\ref{fig:meq-f}, respectively).
The only noticeable difference is in the outer regions where, e.g., \texttt{rh1-f} does not exhibit the fluctuations we described in \texttt{rh1-u}, and shows a global evolution towards EEP.
Nonetheless, even in the filling SCs, $\meq$ stays mostly above $1.0\,\Msun$ (see the blue regions in Fig.~\ref{fig:meq-f}) and drops down to $\gtrsim 0.5$ only before core collapse (see the purple regions in the figure) -- EEP is never reached in the full mass spectrum. Moreover, in the post-core-collapse evolution, the regions where the high-mass stars show temporary EEP based on our fitting (see Appendix~\ref{app:meq}) are far more scattered throughout the cluster than in the underfilling models, and they are most prominent in the \texttt{tan-f} model.

The outer regions of the models \texttt{tan-f}, \texttt{iso-f} and \texttt{rad-f} evolve towards an inverted EEP initially, which is characterised by negative values of $\meq$ (see the red areas in the top row of Fig.~\ref{fig:meq-f}). Within several relaxation times, all three models gradually transition back to positive $\meq$ and none of them is showing an area dominated by $\meq<0$ after core collapse.
The reason behind the inverted EEP is the same here as in the underfilling SCs -- i.e., it is driven by the tangential component of $\disp$ and depends on the number of massive stars with tangential orbits and the outward migration of the low-mass population. Referring to the discussion in Sec.~\ref{sec:meq-u}, a similar comparison can be made between Fig.~\ref{fig:meq-f} (see the extent of the red regions in the bottom row) and Fig.~\ref{fig:disp-f} which shows the difference in velocity dispersion of the high-mass and the low-mass stars in the filling models.

However, the filling models cannot maintain the negative $\meq$ (or the positive difference $\disp[H]-\disp[L]$) for the same amount of time as the underfilling models. First, this is due to the tidal limitation which does not allow the low-mass stars to expand as far as in the underfilling SCs (compare the slopes of the dotted lines in the bottom and top rows of Fig.~\ref{fig:rlLH}). And the second factor is the higher mass-loss rate which prevents the massive stars to stay within their radial shells for a long period of time, even in the tangentially anisotropic model (see the decreasing trend of all uppermost solid black lines in the bottom row of Fig.~\ref{fig:rlLH}).

\section{Conclusions}
\label{sec:concl}

We investigated the implications of the initial stellar kinematics on the global dynamical evolution of star clusters (SCs) with 100k stars. We found that the relaxation processes responsible, e.g., for mass segregation or energy equipartition (EEP) have different timescales in the SCs with tangentially anisotropic, isotropic, or radially anisotropic velocity distributions. We also evaluated the effects of the external Galactic tidal field. Our conclusions are:
\begin{enumerate}
	\item The time of core collapse in all SCs depends on the amount of radial velocity anisotropy introduced to the system. In the tidally underfilling SCs, it happens the earliest in the tangentially anisotropic model and is delayed up to ${\gtrsim}2\,\trh[,0]$ in the radially anisotropic one (the isotropic model is between them). The same order is also visible in the tidally filling SCs, however, the time separation of core collapse is not as pronounced due to the higher mass loss which accelerates the evolution in the radially anisotropic models.
	\item Mass segregation proceeds at different rates in the models \citep[this is consistent with][however, here we extend their conclusions by adding the tangentially anisotropic SC]{pav_ves3}. In particular, (i) in the core region, mass segregation is most rapid in the tangentially anisotropic SC and slows down with the increasing amount of radial anisotropy; (ii) in the outer regions, mass segregation in the tangentially anisotropic SC is the slowest and accelerates in the models with added radial anisotropy.
	\item All tidally underfilling SCs develop radial anisotropy in their outer regions above the half-mass radius whereas the filling SCs form a tangentially anisotropic shell below their tidal radii \citep[this is consistent with][]{Tio_Ves_Var16}.
	\item Focusing on the regions below the half-mass radius, all SCs become isotropic there after several relaxation times (we note that this includes even the SC which started fully tangentially anisotropic). This effect is caused by the much shorter relaxation time scale in the inner regions.
	\item All SCs, regardless of their filling factor, evolve towards EEP in the central regions. The most massive stars $m \gtrsim 0.5\,\Msun$ there are even able to achieve EEP for an extended period of time.
	\item In the outer regions, the evolution differs between the models. The SCs which were initially the most radially anisotropic evolve towards EEP but the other models show an ``inverted'' evolution in the outer regions where the high-mass stars tend to have systematically higher velocity dispersion than the low-mass stars. The duration of this inverted EEP increases with the increasing tangential anisotropy in the system, and decreases as the degree of radial anisotropy in the system grows. This result further extends the findings of \citet{pav_ves_letter,pav_ves2}.
\end{enumerate}

\begin{acknowledgements}
VP has received funding from the European Union's Horizon Europe and the Central Bohemian Region under the Marie Skłodowska-Curie Actions -- COFUND, Grant agreement \href{https://doi.org/10.3030/101081195}{ID~101081195} (``MERIT'').
Views and opinions expressed are, however, those of the authors only and do not necessarily reflect those of the European Union or the Central Bohemian Region. Neither the European Union nor the Central Bohemian Region can be held responsible for them.
VP also acknowledges
(1) the use of the high-performance storage within his project \textit{``Dynamical evolution of star clusters with anisotropic velocity distributions''} at Indiana University Bloomington;
(2) Lilly Endowment, Inc., through its support for the Indiana University Pervasive Technology Institute; 
(3) access to computational resources supplied by the project ``e-Infrastruktura CZ'' (e-INFRA LM2018140) provided within the programme Projects of Large Research, Development and Innovations Infrastructures,
and (4) the support from the project RVO:67985815 at the Czech Academy of Sciences.
ALV acknowledges support from a UKRI Future Leaders Fellowship (MR/S018859/1).
For the purpose of Open Access, the authors have applied a Creative Commons Attribution (CC BY) licence to any Author Accepted Manuscript version arising from this submission.
The \texttt{Python} programming language with \texttt{NumPy} \citep{numpy} and \texttt{Matplotlib} \citep{matplotlib} contributed to this project.
This research has made use of NASA's Astrophysics Data System Bibliographic Services.
\end{acknowledgements}

\bibliographystyle{aa}
\bibliography{bibliography}

\appendix

\section{Plotting the 2D histograms}
\label{app:meq}

In order to plot the very high-resolution Figs.~\ref{fig:aniso-u}, \ref{fig:aniso-f}, \ref{fig:meq-u} \&~\ref{fig:meq-f}, we split the stars from each simulation snapshot into 50 radial shells based on the current Lagrangian radii -- each shell contains 2\,\% of the total SC mass at a given time. We then evaluate $\beta$ or $\meq$ in each shell. The bins without value -- those without stars (e.g., above the tidal radius) or where the simulation terminated earlier are not shown and coloured grey instead.

In the case of the lower rows of Figs.~\ref{fig:aniso-u} \&~\ref{fig:aniso-f} where different mass subgroups are shown, we use the same radial bins as in the top row of each figure (with all stars) but we only evaluate $\beta$ from the stars within the mass range in each shell. This does sometimes result in the shells being empty, hence they are also coloured grey.

To fit the equipartition mass in Figs.~\ref{fig:meq-u} \&~\ref{fig:meq-f}, the stars from each shell are binned by masses with the following bin-edges:
$$\lbrace 0.1, 0.15, 0.2, 0.25, 0.3, 0.35, 0.4, 0.5, 0.6, 0.7, 0.8, 1.0 \rbrace\,\big/\,\Msun$$
Those are arbitrarily selected to account for the IMF generating more low-mass stars and less high-mass stars in the SC.
In each radius--mass bin, we determine $\meq$ using the least-square fitting on Eq.~\eqref{eq:meq}.

Since the velocity dispersion in the central regions of the SCs is subject to high fluctuations (especially during core collapse and the post-core-collapse evolution), the fitted values of $\meq$ in the central mass-bins have high uncertainties -- usually about ${\pm}0.1$ but occasionally also up to ${\pm}0.4$ or more. This especially impacts the regions where $\meq<1$ because the relative error is higher; therefore, in these regions, we only show the fit if its uncertainty is below ${\pm}0.1$ and we colour the other pixels (i.e., with $\meq<1$ and higher uncertainty) grey in the figures. We also note that if the fit is poorly defined due to an insufficient number of data points in the radial-mass bins (as determined by the fitting function) we mask those pixels with grey colour in the figures.

\end{document}